\documentclass[%
 reprint,
 amsmath,amssymb,
 aps,prb
]{revtex4-1}

\usepackage{graphicx}
\usepackage{dcolumn}
\usepackage{bm}
\usepackage{hyperref}
\usepackage{float,epstopdf}

\begin{document}

\preprint{APS/123-QED}

\title{A Complete Analytical Study on the Dynamics of Simple Chaotic Systems}
\author{G. Sivaganesh}
\affiliation{%
Department of Physics, Alagappa Chettiar College of Engineering $\&$ Technology, Karaikudi, Tamilnadu-630 004, India\\
 }%
\author{A. Arulgnanam}
 \email[Corresponding author: ]{gospelin@gmail.com}
\affiliation{%
Department of Physics, St. John's College, Palayamkottai, Tamilnadu-627 002, India\\
 }%
\author{A. N. Seethalakshmi}
\affiliation{%
Department of Physics, The M.D.T hindu College, Tirunelveli, Tamilnadu-627 010, India\\
 }%

\date{\today}

\begin{abstract}
We report in this paper a complete analytical study on the bifurcations and chaotic phenomena observed in certain second-order, non-autonomous, dissipative chaotic systems. One-parameter bifurcation diagrams obtained from the analytical solutions proving the numerically observed chaotic phenomena such as antimonotonicity, period-doubling sequences, Feignbaum remerging have been presented. Further, the analytical solutions are used to obtain the basins of attraction, phase-portraits and Poincare maps for different chaotic systems. Experimentally observed chaotic attractors in some of the systems are presented to confirm the analytical results. The bifurcations and chaotic phenomena studied through explicit analytical solutions is reported in the literature for the first time.

\begin{description}

\item[PACS numbers]
05.45.-a; 05.45.Ac\\
{\bf Keywords:} chaos, antimonotonicity, piecewise-linear
\end{description}
\end{abstract}
                             
\maketitle

\section{Introduction}
\label{sec:1}

Chaos in electronic circuits has been a topic of interest among researchers because of its application to secure communication \cite{Ogorzalek1993,Lakshmanan1994}. On the observation of a chaotic attractor in an autonomous circuit system by Matsumoto \cite{Matsumoto1984}, numerous electronic circuits with chaotic dynamics have been reported \cite{Murali1994,Thamilmaran2000,Arulgnanam2009,Arulgnanam2015}. The implementation of the {\emph{Chua's diode}} using Op-Amps \cite{Kennedy1992} enabled researchers to identify different types of nonlinear elements \cite{Lacy1996,Arulgnanam2009}. Synchronization of chaotic systems has emerged after the study of Pecora and Carroll \cite{Pecora1990} and several electronic circuits with chaotic behavior have also been studied for synchronization \cite{Chua1992,Murali1995}. On the large volume of research in chaos theory only a few have reported analytical studies for chaos and synchronization. The analytical results thus obtained have been used to produce chaotic attractors and synchronization through phase portraits \cite{Thamilmaran2000,Thamilmaran2001,Lakshmanan2003,Arulgnanam2009,Sivaganesh2014,Sivaganesh2015,Arulgnanam2015,Sivaganesh2016a,Sivaganesh2017}. We present in this paper a detailed explicit analytical solution for a class of second-order, non-autonomous chaotic systems with piecewise-linear nonlinear elements. Nonlinear elements with three segmented $voltage-current$ characteristics have been considered for the present study. The dynamics of these of these systems have been studied by presenting one-parameter bifurcation diagrams, phase-portraits, basins of attraction, Poincare maps and power spectrum. A complete explanation of the analytical solutions obtained in each of the piecewise-linear to produce the chaotic attractors in the phase-space is given. \\

\begin{figure}
\begin{center}
\includegraphics[scale=0.4]{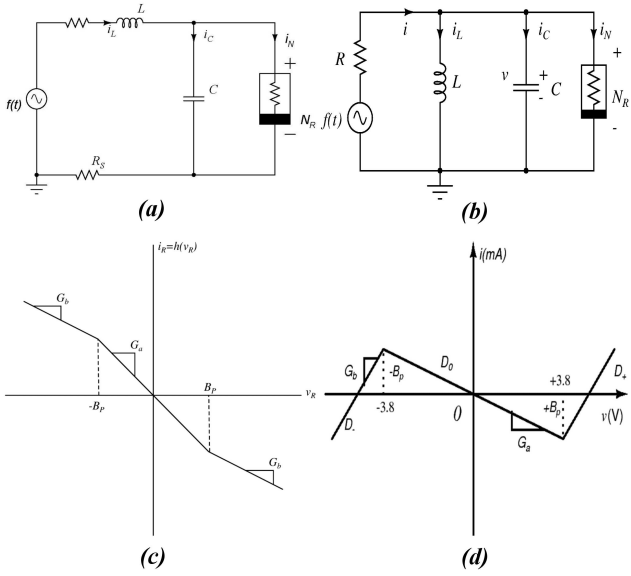}
\caption{(Color online) Schematic representation of the sinusoidally forced (a) series LCR circuit with a nonlinear element $N_R$ connected parallel to the capacitor and (b) parallel LCR circuit with a nonlinear element $N_R$; (c) $V-I$ charactersistics of the {\emph{Chua's diode}} with three negative slope regions and (d)  $V-I$ charactersistics of the {\emph{simplified nonlinear element}} with two positive outer slopes and one negative inner slope.}
\label{fig:1}
\end{center}
\end{figure}

For the present study, we consider two types of second-order circuit systems each with two different nonlinear elements. Fig. \ref{fig:1}(a) and \ref{fig:1}(b) shows the schematic representation of the sinusoidally forced series and parallel LCR circuits with piecewise-linear elements $N_R$. The nonlinear element may be a {\emph{Chua's diode}} \cite{Kennedy1992} or a {\emph{simplified nonlinear element}} \cite{Arulgnanam2009}. The $V-I$ characteristics of the {\emph{Chua's diode}} and the {\emph{simplified nonlinear element}} are as shown in Fig. \ref{fig:1}(c) and Fig. \ref{fig:1}(d), respectively. The paper is divided into two sections. In Section \ref{sec:2} we present the generalized analytical solutions for series LCR circuit systems with piecewise-linear nonlinear elements and present the analytical dynamics of two types of circuit systems. In Section \ref{sec:3},  generalized analytical solutions and analytical dynamics of parallel LCR circuit systems are presented. 

\section{Analytical Dynamics of Series LCR circuit systems}
\label{sec:2}
The circuit equations for a sinusoidally forced series LCR circuit with any three-segmented, piecewise-linear, voltage-controlled nonlinear element is given by
\begin{subequations}
\begin{eqnarray}
C {dv \over dt }  &=&  i_L - g(v), \\
L {di_L \over dt }  &=&  -(R i + R_s) i_L - v + F sin( \Omega t),
\end{eqnarray}
\label{eqn:1}
\end{subequations}
where $g(v)$ is the mathematical form of the piecewise-linear element given by
\begin{equation}
g(v) = G_b v + 0.5(G_b - G_a)[|v+B_p|-|v-B_p|],
\label{eqn:2}
\end{equation}
In terms of the rescaled parameters, the normalized state equations are given as
\begin{subequations}
\begin{eqnarray}
\dot x  &=&  y - g(x), \\ 
\dot y  &=&  -\sigma y - \beta x + f sin(\theta) ,\\ 
\dot \theta  &=&  \omega,
\end{eqnarray}
\label{eqn:3}
\end{subequations}
where, $\sigma = (\beta + \nu \beta)$ and $ \beta = (C/LG^2)$, $ \nu = GR_s$, $ a  = G_a/G$,  $ b = G_b/G$, $f = (f \beta/B_p)$, $\omega = (\omega C/G)$, $ G = 1/R$.
In the piecewise-linear form $g(x)$ can be written as
\begin{equation}
g(x) =
\begin{cases}
bx+(a-b) & \text{if $x \ge 1$}\\
ax & \text{if $|x|\le 1$}\\
bx-(a-b) & \text{if $x \le -1$}
\end{cases}
\label{eqn:4}
\end{equation}
The normalized state equations of the system given by Eq. \ref{eqn:3} is second-ordered in each of the piecewise-linear regions. Hence, an explicit analytical solution can be obtained in each of the  piecewise-linear regions. The generalized analytical solutions for Eq. \ref{eqn:3} is summarized as follows. \\
In the central region $D_0$, $g(x) = ax$ and the dynamical equation of the system can be written as
\begin{equation}
{\ddot y} + {A \dot y} + By = a f sin(\omega t) + f \omega cos(\omega t),
\label{eqn:5}
\end{equation}
where $A = (\sigma+a)$ and $B=(\beta+a \sigma)$. The fixed points in the $D_0$ region is the {\emph{origin}} $(0,0)$. When the roots of the above equation $m_{1,2} = \frac{-A}{2} \pm \frac{\sqrt{A^2-4B}}{2}$ are real and distinct then the state variables $y(t)$ and $x(t)$ are
\begin{subequations}
\begin{eqnarray}
y(t) &=& C_1 e^ {m_1 t} + C_2 e^ {m_2 t} +E_1 +E_2 sin (\omega t) \nonumber \\
&& + E_3 cos (\omega t) \\
x(t) &=& \frac{1}{\beta}(-\sigma y - \dot{y} + f sin (\omega t))
\label{eqn:6}
\end{eqnarray}
\end{subequations}
The constants of the particular integral $E_1, E_2, E_3$ are given as
\begin{subequations}
\begin{eqnarray}
E_1 &=& 0 \\
E_2  &=&  \frac {f  {\omega} ^2 (A-a) + a f B}{A^2 {\omega} ^2 + (B-{\omega} ^2)^2}  \\
E_3  &=&  - \frac {f  \omega (A a +\omega ^2 -B)}{A^2 {\omega} ^2 + (B-{\omega} ^2)^2}
\end{eqnarray}
\label{eqn:7}
\end{subequations}
The constants of the complementary function $C_1$ and $C_2$ are given as
\begin{subequations}
\begin{eqnarray}
C_1 =  &&\frac{e^ {- m_1 t_0}} {m_1 - m_2} \{ ((\sigma - m_2)y_0 - \beta x_0 + m_2 E_1) \nonumber\\
&& + (m_2 E_3 - E_2 \omega) cos \omega t_0 \nonumber \\
&& + ( f+ E_3 \omega + m_2 E_2) sin \omega t_0 \}  \\
C_2 =  &&\frac{e^ {- m_2 t_0}} {m_2 - m_1} \{ ((\sigma - m_1)y_0 - \beta x_0 + m_1 E_1) \nonumber\\
&& + (m_1 E_3 - E_2 \omega) cos \omega t_0 \nonumber \\
&& + ( f+ E_3 \omega + m_1 E_2) sin \omega t_0 \}
\end{eqnarray}
\label{eqn:8}
\end{subequations}
When the roots $m_{1,2}$ are a pair of complex conjugates then 
\begin{subequations}
\begin{eqnarray}
y(t) &=& e^ {ut}(C_1 cos vt + C_2 sin vt) +E_1 +E_2 sin(\omega t) \nonumber \\
&& + E_3 cos(\omega t) \\
x(t) &=& \frac{1}{\beta}(-\sigma y - \dot{y} + f sin \omega t)
\label{eqn:9}
\end{eqnarray}
\end{subequations}
 where $u=\frac{-A}{2}$ and $v=\frac{\sqrt{4B-A^2}}{2}$. The constants $E_1, E_2, E_3$ are the same as given in Eq. \ref{eqn:7} and the constants $C_1, C_2$ are given as 
\begin{subequations}
\begin{eqnarray}
C_1 = && \frac{e^ {- u t_0}} {v} \{((\sigma+u)y_0 +\beta x_0 - u E_1) sin vt_0 \nonumber \\
&& + (y_0-E_1) v cos vt_0 \nonumber \\
&& - ((E_3 \omega + uE_2 + f) sinvt_0 + vE_2 cosvt_0) sin \omega t_0 \nonumber \\
&& + ((E_2 \omega - u E_3) sinvt_0 - v E_3 cosvt_0) cos \omega t_0 \} \\
C_2 = && \frac{e^ {- u t_0}} {v} \{((\sigma+u)y_0 +\beta x_0 - u E_1) cos vt_0 \nonumber \\
&& - (y_0-E_1) v sin vt_0 \nonumber \\
&& - ((E_3 \omega + uE_2 + f) cos vt_0 - vE_2 sin vt_0) sin \omega t_0 \nonumber \\
&& + ((E_2 \omega - u E_3) cos vt_0 + v E_3 sin vt_0) cos \omega t_0 \}
\end{eqnarray}
\label{eqn:10}
\end{subequations}
In the $D_{\pm1}$ region, $g(x) = bx\pm(a-b)$ and the dynamical equation can be written as
\begin{equation}
{\ddot y} + {C \dot y} + Dy = b f sin(\omega t) + f \omega cos(\omega t) \pm \Delta,
\label{eqn:11}
\end{equation}
where $C=(\sigma+b)$, $D=(\beta+b\sigma)$ and $\Delta = \beta(a-b)$. The fixed points corresponding to the $D_{\pm1}$ region are $(\mp \frac{\sigma(a-b)}{(b\sigma+\beta)}, \pm \frac{\beta(a-b)}{(b\sigma+\beta)})$. The state variables $y(t), x(t)$ in this region when the roots $m_{3,4} = \frac{-C}{2} \pm \frac{\sqrt{C^2-4D}}{2}$ are real and distinct is given by
\begin{subequations}
\begin{eqnarray}
y(t) &=& C_3 e^ {m_3 t} + C_4 e^ {m_4 t} +E_4 +E_5 sin (\omega t) \nonumber \\
&& + E_6 cos (\omega t) \\
x(t) &=& \frac{1}{\beta}(-\sigma y - \dot{y} + f sin (\omega t))
\label{eqn:12}
\end{eqnarray}
\end{subequations}
The constants $E_4, E_5, E_6$ are given as 
\begin{subequations}
\begin{eqnarray}
E_4 &=& \pm \frac{\Delta}{D} \\
E_5  &=&  \frac {f  {\omega} ^2 (C-b) + b f D}{C^2 {\omega} ^2 + (D-{\omega} ^2)^2}  \\
E_6  &=&  - \frac {f  \omega (C b +\omega ^2 -D)}{C^2 {\omega} ^2 + (D-{\omega} ^2)^2}
\end{eqnarray}
\label{eqn:13}
\end{subequations}
Here, $+\Delta$ and $-\Delta$ corresponds to $D_{+1}$ and $D_{-1}$ regions, respectively. The constants $C_3,C_4$ are the same as Eq. \ref{eqn:8} except that the constants $a,A,B$ are replaced with $b,C,D$, respectively. When the roots $m_{3,4}$ are a pair of complex conjugates, the state variables are given as
\begin{subequations}
\begin{eqnarray}
y(t) &=& e^ {ut}(C_3 cos vt + C_4 sin vt) +E_4 +E_5 sin(\omega t) \nonumber \\
&& + E_6 cos(\omega t) \\
x(t) &=& \frac{1}{\beta}(-\sigma y - \dot{y} + f sin \omega t)
\label{eqn:14}
\end{eqnarray}
\end{subequations}
where $u=\frac{-C}{2}$ and $v=\frac{\sqrt{4D-C^2}}{2}$. The constants $C_3,C_4$ are the same as Eq. \ref{eqn:10} except that the constants $a,A,B$ are replaced with $b,C,D$, respectively. The solutions presented above for each piecewise-linear regions can be used to obtain the trajectory of the system in the corresponding region for a given initial condition $(x_0,y_0)$. The state variables $(x,y)$ obtained every instant of time acts as the initial condition for obtaining the state variables in the next instant. The state variables obtained in the individual regions can be plotted together to produce the complete phase-space trajectory. Now, we present the dynamics of series LCR circuit systems with two different types of nonlinear elements, using the analytical solutions obtained above. \\

\begin{figure}
\begin{center}
\includegraphics[scale=0.4]{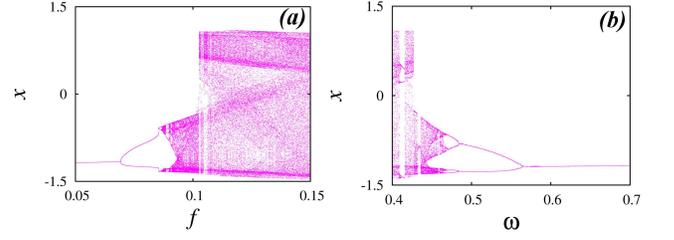}
\caption{(Color online) Analytically obtained one-parameter bifurcation diagram of the MLC circuit: (a) {\emph{amplitude scanning}} in the $f-x$ plane with a fixed value of $\omega=0.72$ and (b) {\emph{frequency scanning}} in the $\omega -x$ plane with the amplitude fixed at $f=0.053$.}
\label{fig:2}
\end{center}
\end{figure}

\begin{figure}
\begin{center}
\includegraphics[scale=0.4]{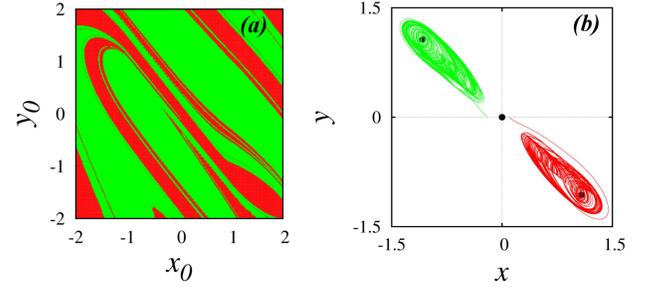}
\caption{(Color online) (a) Analytically obtained basins of attraction for the one-band chaotic attractor in the $(x_0-y_0)$ phase-plane for the MLC circuit. The regions are color coded as follows: green-chaotic attractor corresponding to the left half plane and red- chaotic attractor corresponding to the right half plane. (b) One-band chaotic attractors originating from different initial conditions obtained from the basin of attraction near the fixed point $(0,0)$.}
\label{fig:3}
\end{center}
\end{figure}

\begin{figure}
\begin{center}
\includegraphics[scale=0.4]{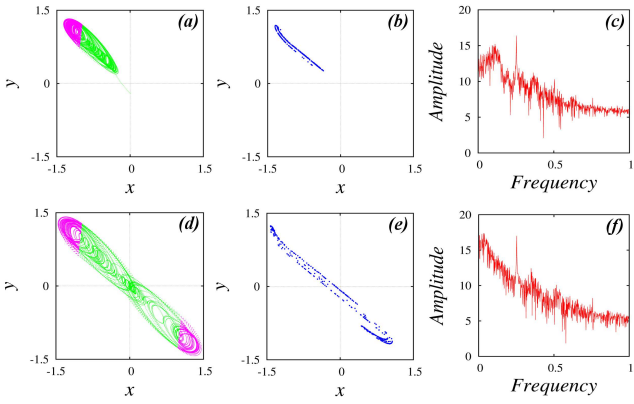}
\caption{(Color online) Analytically obtained  (a) One-band chaotic attractor in each of the piecewise linear regions $D_0$ (green) and $D_{+1}$ (magenta) in the $x-y$ phase plane for the initial conditions $(x_0,y_0 = 0,-0.2)$ and its corresponding (b) Poincare map and (c) Power spectrum; Analytically obtained (d) double-band chaotic attractor in the the piecewise-linear regions $D_0$ (green) and $D_{\pm1}$ (magenta) in $x-y$ phase plane and its corresponding (e) Poincare map and (f) Power spectrum.}
\label{fig:4}
\end{center}
\end{figure}

\subsection{Murali-Lakshmanan-Chua Circuit}

The {\emph{Murali-Lakshmanan-Chua (MLC)}} circuit is a series LCR circuit with a {\emph{Chua's diode}} connected parallel to the capacitor. The circuit exhibits a wide range of chaotic behavior in its dynamics. It has been studied for chaotic, strange non-chaotic and synchronization behaviors for the past two decades \cite{Murali1994,Murali1995,Sivaganesh2014,Sivaganesh2015}. This circuit exhibits two prominent chaotic attractors at the control parameter values $f =0.1$ and $0.14$, respectively. The one-band chaotic attractor is obtained through a period-doubling route as shown in the one-parameter bifurcation diagram obtained analytically in Fig. \ref{fig:2}(a). The chaotic dynamics of the circuit exhibiting a reverse period-doubling route with the frequency of the external force as control parameter is shown in Fig. \ref{fig:2}(b). The analytical solutions can be used to obtain the basins of attraction corresponding to one-band chaotic attractors. Fig. \ref{fig:3}(a) shows the basin of attractor corresponding to the one-band chaotic attractor in the $(x_0-y_0)$ phase space. The green colored regions indicate the set of initial conditions that settle down at the one-band chaotic attractor in the left half plane of the phase space while the red colored regions indicate the attractor corresponding to the right half plane, respectively. Fig. \ref{fig:3}(b) shows the one-band chaotic attractors originating from their corresponding colored basins shown in Fig. \ref{fig:3}(a).  The fixed points (black dots) in each of the piecewise-linear regions shows that they form an attracting set in the phase-space around which the one-band chaotic attractors settles down asymptotically.  The analytical solutions can further be used to obtain phase-portraits and Poincare maps for the chaotic attractors as shown in Fig. \ref{fig:4}. The state variables $x(t)$ and $y(t)$ obtained in each of the piecewise-linear regions are plotted to produce the chaotic attractors. The one-band chaotic attractor shown in Fig. \ref{fig:4}(a) exists only in the $D_{-1}$ and $D_0$ piecewise-linear regions. The Poincare maps and the power spectra indicating a broader range of frequency distribution corresponding to the one-band chaotic attractor are shown in Fig. \ref{fig:4}(b) and \ref{fig:4}(b), respectively. The phase portrait, Poincare map and power spectrum corresponding to the double band chaotic attractor are given in Fig. \ref{fig:4}(d)-\ref{fig:4}(f). 


\begin{figure}
\begin{center}
\includegraphics[scale=0.4]{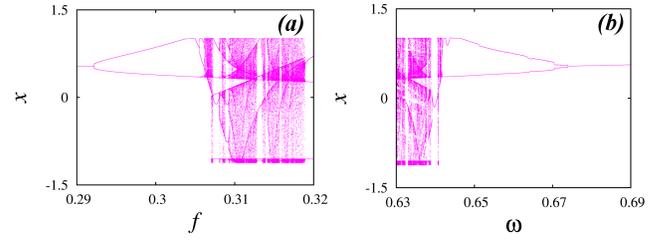}
\caption{(Color online) Analytically obtained one-parameter bifurcation diagram of the forced series LCR circuit with a {\emph{simplified nonlinear element}}: (a) {\emph{amplitude scanning}} in the $f-x$ plane with a fixed value of $\omega=0.7084$ indicating the period-doubling route to chaos and (b) {\emph{frequency scanning}} in the $\omega -x$ plane with the amplitude fixed at $f=0.28$.}
\label{fig:6}
\end{center}
\end{figure}

\begin{figure}
\begin{center}
\includegraphics[scale=0.4]{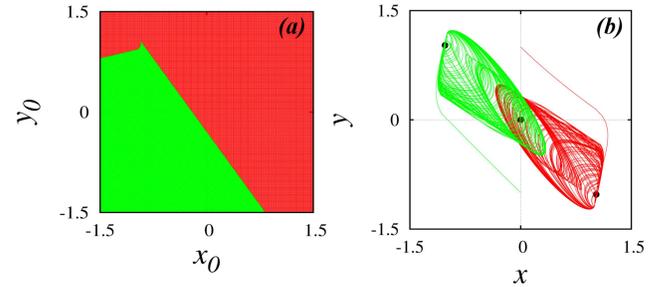}
\caption{(Color online) (a) Analytically obtained basins of attraction for the one-band chaotic attractor in the $(x_0-y_0)$ phase-plane for the series LCR circuit with a {\emph{simplified nonlinear element}}. The regions are color coded as follows: green-chaotic attractor corresponding to the left half plane and red- chaotic attractor corresponding to the right half plane. (b) One-band chaotic attractors originating from different initial conditions obtained from the basin of attraction.}
\label{fig:7}
\end{center}
\end{figure}

\begin{figure}
\begin{center}
\includegraphics[scale=0.4]{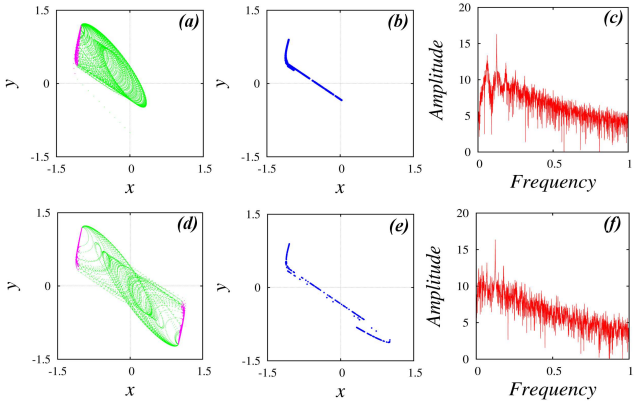}
\caption{(Color online) Analytically obtained  (a) One-band chaotic attractor in each of the piecewise linear regions $D_0$ (green) and $D_{+1}$ (magenta) in the $x-y$ phase plane for the initial conditions $(x_0,y_0 = 0,-1)$ and its corresponding (b) Poincare map and (c) Power spectrum; Analytically obtained (d) double-band chaotic attractor in the the piecewise-linear regions $D_0$ (green) and $D_{\pm1}$ (magenta) in $x-y$ phase plane and its corresponding (e) Poincare map and (f) Power spectrum.}
\label{fig:8}
\end{center}
\end{figure}

\subsection{Forced series LCR circuit with a {\emph{Simplified nonlinear element}}}

This circuit introduced by Arulgnanam {\emph{et al}} \cite{Arulgnanam2009} produces chaotic attractors with a least number of circuit elements. Further, the fractal dimension of the chaotic attractors observed in this system is found to have a larger value as compared to other second-order chaotic systems. Figure \ref{fig:5} shows the experimentally observed chaotic attractor in the $(v-i_L)$ phase-plane for the circuit parameters $C=10.32$ nF, $L=42.6$ mH, $R=2050~ \Omega$, $G_a = -0.56$ mS, $G_b = +2.5$ mS, $B_p = \pm 1.0$ V and $F=4.59$ V. The circuit also exhibits a wide range of chaotic behavior similar to the MLC circuit. The analytically obtained one-parameter bifurcation diagram in the $f-x$ and $\omega-x$ planes indicating the period-doubling and reverse period-doubling sequences observed in the circuit dynamics is shown in Fig. \ref{fig:6}(a) and \ref{fig:6}(b), respectively. The analytically obtained basins of attraction for the one-band chaotic attractors shown in Fig. \ref{fig:7}(a) indicates the set of initial conditions that settles down into a one-band chaotic attractor at the right half plane (red) and the left half plane. The one-band chaotic attractors arising from two different colored basins shown in Fig. \ref{fig:7}(a) along with the fixed points (black dots) of the three piecewise-linear regions is shown in Fig. \ref{fig:7}(b). The analytically observed one-band chaotic attractor in the piecewise-linear regions and its corresponding Poincare map and power spectrum obtained at the amplitude of the external force $f=0.3065$ is shown in Fig. \ref{fig:8}(a)-\ref{fig:8}(c). Fig. \ref{fig:8}(d)-\ref{fig:8}(f) shows the double-band chaotic attractor and its corresponding Poincare map and power spectrum obtained at the value of amplitude $f=0.31$. 

\section{Analytical Dynamics of Parallel LCR circuit systems}
\label{sec:3}

The state equations of a forced parallel LCR circuit system with a three-segmented piecewise nonlinear element connected parallel to the capacitor is given as s 
\begin{subequations}
\begin{eqnarray} 
C {dv \over dt } & = & {1\over R} (F sin( \Omega t)- v)- i_L - g(v), \\
L {di_L \over dt } & = & v,
\end{eqnarray}
\label{eqn:15}
\end{subequations}
The piecewise-linear function $g(v)$ is as given in Eq. \ref{eqn:2}. After proper rescaling, the normalized state equations of the system can be written as
\begin{subequations}
\begin{eqnarray}
\dot x & = & f sin(\theta) - x - y - g(x), \\
\dot y & = & \beta x , \\
\dot \theta & = & \omega,
\end{eqnarray}
\label{eqn:16}
\end{subequations}
where $x = v/B_p$, $y = (i_L / G B_p)$,  $ \beta = (C/LG^2)$, $ a  = G_a/G$,  $ b = G_b/G$, $f = (F \beta/B_p)$, $\omega = (\Omega C/G)$ and  $ G = 1/R$. The mathematical form of the function $g(x)$ is as given in Eq. \ref{eqn:4}. The explicit analytical solutions obtained for the normalized state equations given in Eq. \ref{eqn:16} is summarized as follows.\\
In the central region $D_0$, $g(x)=ax$ and the dynamical equations of the system can be written as
\begin{equation}
{\ddot y} + {A \dot y} + By = f \beta sin(\omega t),
\label{eqn:17}
\end{equation}
where $A=1+a$ and $B=\beta$. The {\emph{origin}} $(0,0)$ acts as the fixed point in this region. The state variables of the system when the roots ${m_{1,2}} =  \frac{-(A) \pm \sqrt{(A^{2}-4B)}} {2}$ are real and distinct is given by
\begin{subequations}
\begin{eqnarray}
y(t) &=& C_1 e^ {m_1 t} + C_2 e^ {m_2 t} + E_1 + E_2 sin (\omega t) \nonumber \\
&&+ E_3 cos (\omega t) \\
x(t) &=& \frac{1}{\beta}(\dot{y}), 
\end{eqnarray}
\label{eqn:18}
\end{subequations}
The constants $E_1,E_2,E_3$ are given as
\begin{subequations}
\begin{eqnarray}
E_1 &=& 0 \\
E_2  &=&  \frac {f\beta(B-\omega^2)}{A^2 {\omega} ^2 + (B-{\omega} ^2)^2}  \\
E_3  &=&  - \frac {f A \beta \omega}{A^2 {\omega} ^2 + (B-{\omega} ^2)^2}
\end{eqnarray}
\label{eqn:19}
\end{subequations}
The constants $C_1, C_2$ are given as
\begin{subequations}
\begin{eqnarray}
C_1 =  &&\frac{e^ {- m_1 t_0}} {m_1 - m_2} \{ (\beta x_0 - m_2 y_0+m_2 E_1) \nonumber \\
&& + (m_2 E_3 - \omega E_2) cos \omega t_0 \nonumber \\
&& + (\omega E_3 + m_2 E_2) sin \omega t_0 \}  \\
C_2 =  &&\frac{e^ {- m_2 t_0}} {m_2 - m_1} \{ (\beta x_0 - m_1 y_0 + m_1 E_1)  \nonumber \\
&& + (m_1 E_3 - \omega E_2) cos \omega t_0 \nonumber \\
&& + (\omega E_3 + m_1 E_2) sin \omega t_0 \} 
\end{eqnarray}
\label{eqn:20}
\end{subequations}
When the roots are a pair of complex conjugates, then the state variables are 
\begin{subequations}
\begin{eqnarray}
y(t) &=& e^ {ut}(C_1 cos vt + C_2 sin vt) + E_1 + E_2 sin \omega t \nonumber \\ 
&&+ E_3 cos \omega t,\\
x(t) &=& \frac{1}{\beta}(\dot{y}), 
\end{eqnarray}
\label{eqn:21}
\end{subequations}
where, $u=\frac{-A}{2}$ and $v=\frac{\sqrt(4B-A^{2})}{2}$. The constants $C_1, C_2$ are given as
\begin{subequations}
\begin{eqnarray}
C_1 = && -\frac{e^ {- u t_0}} {v} \{((\beta x_0 - u y_0 + u E_1) sin vt_0 - v y_0 cos vt_0 \nonumber \\
&& - ((E_2 \omega - uE_3) sinvt_0 - v E_3 cosvt_0) cos \omega t_0 \nonumber \\
&& + ((E_3 \omega + u E_2) sinvt_0 + v E_2 cosvt_0) sin \omega t_0 \} \nonumber \\ \\
C_2 = && -\frac{e^ {- u t_0}} {v} \{((\beta x_0 - u y_0 + u E_1) cos vt_0 + v y_0 sin vt_0 \nonumber \\
&& - ((E_2 \omega - uE_3) cos vt_0 + v E_3 sin vt_0) cos \omega t_0 \nonumber \\
&& + ((E_3 \omega + u E_2) cos vt_0 - v E_2 sin vt_0) sin \omega t_0 \} \nonumber \\
\end{eqnarray}
\label{eqn:22}
\end{subequations}
In the $D_{\pm 1}$ regions, $g(x)=bx \pm (a-b)$ and the dynamical equation of the system can be written as
\begin{equation}
{\ddot y} + {C \dot y} + Dy = f \beta sin(\omega t)  \mp \Delta,
\label{eqn:23}
\end{equation}
where $C=(1+b)$, $D=\beta$ and $-\Delta$, $+\Delta$ corresponds to the $D_{+1}$ and $D_{-1}$ regions, respectively. The fixed points in the $D_{\pm1}$ regions are $(0, \pm b-a)$.  The state variables in these regions when the roots  $m_{3,4} = \frac{-C}{2} \pm \frac{\sqrt{C^2-4D}}{2}$ are real and distinct are given as
\begin{subequations}
\begin{eqnarray}
y(t) &=& C_3 e^ {m_3 t} + C_4 e^ {m_4 t} + E_4 + E_5 \sin \omega t  \nonumber \\ 
&&+ E_6 \cos \omega t {\pm} \Delta,\\
x(t) &=& \frac{1}{\beta}(\dot{y}), 
\end{eqnarray}
\label{eqn:24}
\end{subequations}
The constants $E_3, E_4, E_5$ are given as 
\begin{subequations}
\begin{eqnarray}
E_4 &=& \mp \Delta \\
E_5 &=&  \frac {f \beta(D-\omega^2)}{C^2 {\omega} ^2 + (D-{\omega} ^2)^2}  \\
E_6  &=&  - \frac {D f \beta \omega}{C^2 {\omega} ^2 + (D-{\omega} ^2)^2}
\end{eqnarray}
\label{eqn:25}
\end{subequations}
The constants $C_3,C_4$ are the same as Eq. \ref{eqn:20} except that the constants $a,A,B$ are replaced with $b,C,D$, respectively. When the roots $m_{3,4}$ are a pair of complex conjugates, the state variables are given as
\begin{subequations}
\begin{eqnarray}
y(t) &=& e^ {ut}(C_3 cos vt + C_4 sin vt) + E_4 + E_5 sin(\omega t) \nonumber \\ 
&&+ E_6 cos(\omega t) {\pm} \Delta,\\
x(t) &=& \frac{1}{\beta}(\dot{y}), 
\end{eqnarray}
\label{eqn:26}
\end{subequations}
where $u=\frac{-C}{2}$ and $v=\frac{\sqrt{4D-C^2}}{2}$. The constants $C_3,C_4$ are the same as Eq. \ref{eqn:22} except that the constants $a,A,B$ are replaced with $b,C,D$, respectively. Now, we discuss the analytical dynamics of the parallel LCR circuit systems with the {\emph{Chua's diode}} and the {\emph{simplified nonlinear element}} as the nonlinear elements.

\subsection{Variant of Murali-Lakshmanan-Chua Circuit}

The {\emph{Variant of Murali-Lakshmanan-Chua (MLCV)}} circuit introduced by Thamilmaran {\emph{et al}} \cite{Thamilmaran2000} presents a rich variety of bifurcations and chaos in its dynamics such as the quasiperiodic, reverse period-doubling routes to chaos, antimonotonicity and {\emph{remerging Feignbaum trees}} to name a few \cite{Thamilmaran2001}. Using the analytical solutions presented above we discuss some of the bifurcation and chaotic phenomena observed in the circuit dynamics. The analytically obtained one-parameter bifurcation in the $f-x$ plane shown in Fig. \ref{fig:9}(a) reveals the antimonotonicity behavior observed in the dynamics of the circuit. The period-doubling route to chaos results in a reverse period-doubling sequence with the increase in the amplitude of the external force $f$. Fig. \ref{fig:9}(b) shows the one-parameter bifurcation diagram obtained in the $\omega-x$ plane. Another interesting phenomena named the {\emph{remerging Feignbaum trees}} observed numerically in the circuit dynamics\cite{Thamilmaran2001}, is explained using the analytically obtained one-parameter bifurcation diagrams shown in Fig. \ref{fig:10}. Further, a plot of the analytical solutions of each piecewise-linear regions results in chaotic attractors shown in Fig. \ref{fig:11}(a) and \ref{fig:11}(d) for the amplitudes $f=0.375$ and $f=0.411$, respectively. The chaotic attractors are settle down in a region of space around the unstable and stable fixed points (black dots) at the $D_0$ and $D_{\pm1}$ regions, respectively. The Poincare map and the power spectrum of the chaotic attractors shown in  Fig. \ref{fig:11}(a) and \ref{fig:11}(d) are presented in Fig. \ref{fig:11}(b), \ref{fig:11}(c) and Fig. \ref{fig:11}(e), \ref{fig:11}(f), respectively.

\subsection{Forced parallel LCR circuit with a {\emph{Simplified nonlinear element}}}

The forced parallel LCR circuit with a {\emph{simplified nonlinear element}} introduced by Arulgnanam {\emph{et al}} \cite{Arulgnanam2015} exhibits a torus breakdown and reverse period-doubling routes to chaos. The circuit exhibits two prominent chaotic attractors at the amplitudes of the external force $F=4.75$ and $F=4.779$ with other circuit parameters fixed at 
$C=13.13$ nF, $L=163.6$ mH, $R=2.05~k \Omega$, $G_{a} = -0.56$ mS, $G_{b} = +2.5$ mS and $B_{p} = \pm~3.8$ V, respectively. The rescaled parameters of the circuit are $\beta = 0.2592,~a=-1.148,~b=5.125$ and $\nu = 1.421$ kHz. The experimentally observed chaotic attractors and their corresponding analytically observed power spectra are shown in Fig. \ref{fig:12}. The entire dynamics of the circuit observed analytically through one-parameter bifurcation diagrams in the $(f-x)$ and $(\omega-x)$ planes are shown in Fig. \ref{fig:13}(a) and \ref{fig:13}(b), respectively. From Fig. \ref{fig:13} we could observe that the circuit exhibits chaotic behavior over a wide range of the amplitude and frequency of the external force. Fig. \ref{fig:14}(a) and \ref{fig:14}(d) shows the analytically obtained chaotic attractors plotted in each of the piecewise-linear regions $D_0$ (red), $D_{\pm1}$ (green) for the amplitudes $f=0.695$ and $f=0.855$, respectively. The Poincare map and the power spectrum of the chaotic attractors shown in  Fig. \ref{fig:14}(a) and \ref{fig:14}(d) are presented in Fig. \ref{fig:14}(b), \ref{fig:14}(c) and Fig. \ref{fig:14}(e), \ref{fig:14}(f), respectively.

\section{Conclusions}

In this paper we have reported the effective application of an analytical solution for identification of several interesting phenomena in simple chaotic systems. The antimonotonicity, period-doubling, reverse period-doubling and Feignbaum remerging identified earlier through numerical studies have been proved analytically. Phase-portraits revealing the existence of trajectories in individual piecewise-linear regions have been presented. The efficiency of this solution can be applied for analytically studying five or more-segmented piecewise-linear element second-order chaotic circuit systems. Explicit analytical solutions of this kind paves way for a better understanding on the chaotic phenomena observed in simple circuit systems.

\section*{Acknowledgements}

One of the authors A. Arulgnanam gratefully acknowledges Dr.K. Thamilmaran, Centre for Nonlinear Dynamics, Bharathidasan University, Tiruchirapalli, for his help and permission to carry out the experimental work during his doctoral programme.

\end{document}